%% file: paper.final.tex
\documentclass[copyright,creativecommons]{eptcs}

\providecommand{\event}{PLACES 2019} 

\usepackage{underscore}           

\usepackage{csquotes}
\usepackage{fancyvrb}
\usepackage{booktabs}
\usepackage{adjustbox}
\usepackage{xspace}
\usepackage{relsize}
\usepackage[nomain,
            order=word,
            hyperfirst=false,
            acronym,
            shortcuts,
            nonumberlist]{glossaries}

\usepackage{listings}
\usepackage{subcaption}
\usepackage{cleveref}
\usepackage{comment}
\usepackage{idrislang}
\usepackage{colour-blind}
\usepackage{inconsolata}

\DefineVerbatimEnvironment%
  {FigureVerbatim}
  {BVerbatim}
  {commandchars=\\\{\},fontsize=\smaller}

\DefineVerbatimEnvironment%
  {InlineVerbatim}
  {Verbatim}
  {commandchars=\\\{\},xleftmargin=\parindent}

\newcommand{\furl}[1]{\footnote{\url{#1}}}
\loadglsentries[type=\acronymtype]{./acronyms.gloss}%
\glsdisablehyper{}%
\makeglossaries{}

\makeatletter
\DeclareRobustCommand\onedot{\futurelet\@let@token\@onedot}
\def\@onedot{\ifx\@let@token.\else.\null\fi\xspace}

\def\ie{{i.e}\onedot}
\def\cf{{cf}\onedot}

\makeatother

\newcommand{\naive}{na{\"i}ve}
\newcommand{\Naive}{Na{\"i}ve}

\graphicspath{{image/}{figure/}{logo/}}
\DeclareGraphicsExtensions{.pdf,.png,.jpg,.eps}

\fvset{commandchars=\\\{\}}

\title{Value-Dependent Session Design in a Dependently Typed Language}

\author{%
  Jan de Muijnck-Hughes%
  \qquad%
  Wim Vanderbauwhede%
  \institute{University of Glasgow}%
  \email{Jan.deMuijnck-Hughes@glasgow.ac.uk}%
  \email{Wim.Vanderbauwhede@glasgow.ac.uk}%
  \and%
  Edwin Brady%
  \institute{University of St Andrews}%
  \email{ecb10@st-andrews.ac.uk}%
}

\def\titlerunning{Value-Dependent Session Design in a Dependently Typed Language}
\def\authorrunning{Jan de Muijnck-Hughes et al.\@}

\begin{document}
\maketitle{}

\begin{abstract}
  Session Types offer a typing discipline that allows protocol specifications to be used during type-checking, ensuring that implementations adhere to a given specification.
  When looking to realise global session types in a dependently typed language care must be taken that values introduced in the description are used by roles that know about the value.

We present \texttt{Sessions}, a Resource Dependent \ac{edsl} for describing global session descriptions in the dependently typed language Idris.
As we construct session descriptions the values parameterising the \acp{edsl}' type keeps track of roles and messages they have encountered.
We can use this knowledge to ensure that message values are only used by those who know the value.
\texttt{Sessions} supports protocol descriptions that are computable, composable, higher-order, and value-dependent.
We demonstrate \texttt{Sessions} expressiveness by describing the TCP Handshake, a multi-modal server providing echo and basic arithmetic operations, and a Higher-Order protocol that supports an authentication interaction step.
\end{abstract}

\input{./content/introduction.tex}
\input{./content/language.tex}
\input{./content/examples.tex}
\input{./content/discussion.tex}
\input{./content/related.tex}
\input{./content/conclusion.tex}

\bibliographystyle{eptcs}
\bibliography{biblio}

\end{document}

%% file: content/introduction.tex
\section{Introduction}\label{sec:introduction}

\Acp{mpst} are a well-known typing discipline for describing sequences of message exchanges over communication channels~\cite{DBLP:journals/jacm/HondaYC16}.
Global session types present an overview of the interactions made between each component, and local session types describe the interactions respective to each participant.
When realising \acp{mpst} both global and local types are commonly presented either as bespoke
\acp{dsl}~\cite{Yoshida2014spl}; \acp{edsl} that extend an existing programming language~\cite{Kouzapas2016tpm}; or a language extension~\cite{Lindley2017lfs}.

\acp{mpst} are, however, limited in their expressiveness.
Session designers cannot specify value dependencies between sent messages, or reason about message contents.
Recent work has extended the theory of \ac{mpst} to incorporate value based reasoning on messages~\cite{Toninho2016,DBLP:conf/fossacs/ToninhoY18,DBLP:conf/concur/BocchiHTY10,DBLP:conf/cc/NeykovaHYA18}.

Dependent types allow for more precise reasoning on programs by allowing types to depend on values.
Existing work has shown how dependent types provide greater control over programs, and how they can be used to reason about communication in concurrent programs~\cite{DBLP:journals/fuin/BradyH10,DBLP:journals/aghcs/Brady17,Brady2016tdd,DBLP:conf/fossacs/ToninhoY18} and communicating systems~\cite{Guenot2015}.
More generally, earlier work~\cite{Brady2014rda,Brady2012rsp} has shown how dependent types can provide resource-based type-level reasoning about programs.

\begin{figure}[ht]
  \centering
  \begin{subfigure}{0.475\textwidth}
    \input{./figure/naive/specification}
  \caption{\label{fig:intro:naive:def}Specification.}
  \end{subfigure}
  \begin{subfigure}{0.475\textwidth}
    \input{./figure/naive/example}
    \caption{\label{fig:intro:naive:example}Example}
  \end{subfigure}
  \caption{\label{fig:intro:naive}\Naive{} Realisation of an \glsentryshort{edsl} for describing Global Session Types.}
\end{figure}

Idris is a general purpose functional language with dependent types~\cite{Brady2013igp},
\Cref{fig:intro:naive:def} presents a \naive{} implementation of a global session description as an \ac{edsl} within Idris.
The type for expressions, \IdrisType{Session}, is indexed by a type describing participants (\IdrisBound{role}) and type associated with expressions---\IdrisBound{type}.
Message passing is described using \IdrisData{Send} in which the sending and receiving roles are specified, together with the type of message being sent.
Expressions are sequenced using a standard \IdrisData{Let} construct.
This is possible by indexing \IdrisType{Session} by the type associated with individual expressions.
The \IdrisData{Let} expression also leads to Do-notation through implementation of Idris' monadic bind operator---\IdrisFunction{(>>=)}.
Sessions descriptions are concluded using \IdrisData{End}.

\Cref{fig:intro:naive:example} demonstrates use of \IdrisType{Session} to describe the well-known \texttt{TCP}-handshake~\cite{rfc793}.
The roles of \enquote{Alice} and \enquote{Bob} describe the client and server, and \IdrisType{Packet} describes the \texttt{TCP} messages themselves.
Each send operation describes how Alice and Bob send the packet and sequence numbers in a tuple.
However, our session description does not describe the value dependencies between the sequence numbers, nor which constructor of \IdrisType{Packet} should be sent.
Specifically, we should be able to provide guarantees that the sequence numbers are incremented by one when they are returned to their originators.

\begin{figure}[ht]
  \centering
  \begin{subfigure}{1.0\textwidth}
    \input{./figure/naive/specification.value}
    \caption{\label{fig:intro:naive:example:value}Value Dependent Specification}
  \end{subfigure}

  \begin{subfigure}{1.0\textwidth}
    \input{./figure/naive/specification.error}
    \caption{\label{fig:intro:naive:example:error}Errorful Value Dependent Specification}
  \end{subfigure}
  \caption{\label{fig:intro:naive:examples}Global Session Types described using \ac{edsl} presented in \Cref{fig:intro:naive:def}.}
\end{figure}

The \IdrisData{Let} construct of \IdrisType{Session} allows binding of, and pattern matching on, message values to an identifier that can be reasoned about using standard Idris constructs.
\Cref{fig:intro:naive:example:value} demonstrates how we can use these \emph{bound} values to describe messages that rely on previously seen values.
A dependent pair provides existential quantification that the type of Bob's response to Alice contains the sequence number \IdrisBound{x} incremented by one.
Standard equality types \IdrisType{(=)} ensure this relation is available at the type-level.
Similar use of dependent pairs ensures that the same incremented sequence number is returned to Bob, together with their sequence number \IdrisBound{y} also incremented by one.
However, our \ac{edsl} does not reason about message origin, nor who is aware of these values.
\Cref{fig:intro:naive:example:error} demonstrates how we can insert a third participant \enquote{Charlie} into the protocol who can send a message to Alice whose type depends on a value that Charlie is not aware of.
When specifying \acp{edsl} for global session descriptions in a dependently typed language we must prevent roles from depending on values they are not aware of.

\subsection{Contributions}\label{sec:contribution}

Taking inspiration from \aclp{mpst} we have been investigating how dependent types, as presented in Idris, can be used to design, implement, and reason about communicating systems.
Existing work has presented a type-driven approach for designing and developing communicating concurrent systems~\cite{DBLP:journals/aghcs/Brady17}.
A limitation in this work is not being able to specify value-dependencies on messages for multi-party protocols.
This paper presents the initial outcomes of our research to address this limitation.

Resource-Dependent \acp{edsl} are a common design pattern associated with \ac{edsl} construction in Idris~\cite{Brady2016tdd,Brady2016sms} derived from existing work on parameterised monads and Hoare monads~\cite{Atkey2009pnc,Borgstrom2011vsp}.
Such construction allows us to associate, and manage, an abstract state within the type of the \ac{edsl} itself.
Using this construction we can index the type of our \ac{edsl} by a \emph{knowledge index} that captures the messages seen by each participant.
Using this index we ensure that value dependent messages are only sent by participants using values that the participant knows about.

Our main contributions are:

\begin{enumerate}
\item
  \texttt{Sessions}, a Resource-Dependent \ac{edsl} to specify global session descriptions.
  \texttt{Sessions} improves upon existing \ac{edsl} designs by introducing global session descriptions that are computable, composable, higher-order, and value-dependent.

\item
  Several example session descriptions that describe the TCP Handshake (\Cref{sec:example:tcp}), a multi-modal server providing echo and basic arithmetic operations (\Cref{sec:example:server}), and a higher-order session description that emulates an authentication interaction---\Cref{sec:example:hoppy}.
\end{enumerate}

\texttt{Sessions} has been realised using Idris, however, the construction techniques presented are agnostic to dependently typed languages.
Other languages such as Agda~\cite{Norell2009agda} should be capable of realising this \ac{edsl}.


%% file: figure/naive/specification.tex
\begin{FigureVerbatim}
\IdrisKeyword{data} \IdrisType{Session} : (\IdrisBound{type}, \IdrisBound{role} : \IdrisType{Type})
            -> \IdrisType{Type} \IdrisKeyword{where}
  \IdrisData{Let}  : (\IdrisBound{this} : \IdrisType{Session} \IdrisBound{a} \IdrisBound{role})
      -> (\IdrisBound{into} : \IdrisBound{a} -> \IdrisType{Session} \IdrisBound{b} \IdrisBound{role})
      -> \IdrisType{Session} \IdrisBound{b} \IdrisBound{role}

  \IdrisData{Send} : (\IdrisBound{from}, \IdrisBound{to} : \IdrisBound{role})
      -> (\IdrisBound{type} : \IdrisType{Type})
      -> \IdrisType{Session} \IdrisBound{type} \IdrisBound{role}
  \IdrisData{End}  : \IdrisType{Session} \IdrisType{()} \IdrisBound{role}
\end{FigureVerbatim}


%% file: figure/naive/example.tex
\begin{FigureVerbatim}
\IdrisKeyword{data }\IdrisType{Packet} = \IdrisData{Syn} | \IdrisData{SynAck} | \IdrisData{Ack}

\IdrisFunction{tcpHandshakeN} : (\IdrisBound{alice},\IdrisBound{bob} : \IdrisBound{role})
             -> \IdrisType{Session} \IdrisType{()} \IdrisBound{role}
\IdrisFunction{tcpHandshakeN} \IdrisBound{alice} \IdrisBound{bob} = \IdrisKeyword{do}
   \IdrisData{Send} \IdrisBound{alice} \IdrisBound{bob} \IdrisType{(\IdrisType{Packet\IdrisType{,}}} \IdrisType{Nat\IdrisType{)}}
   \IdrisData{Send} \IdrisBound{bob} \IdrisBound{alice} \IdrisType{(\IdrisType{Packet\IdrisType{,}}} \IdrisType{Nat\IdrisType{,}} \IdrisType{Nat\IdrisType{)}}
   \IdrisData{Send} \IdrisBound{alice} \IdrisBound{bob} \IdrisType{(\IdrisType{Packet\IdrisType{,}}} \IdrisType{Nat\IdrisType{,}} \IdrisType{Nat\IdrisType{)}}
   \IdrisData{End}
\end{FigureVerbatim}


%% file: figure/naive/specification.value.tex
\begin{FigureVerbatim}
\IdrisFunction{tcpHandshakeDep} : (\IdrisBound{alice},\IdrisBound{bob} : \IdrisBound{role}) -> \IdrisType{Session} \IdrisType{()} \IdrisBound{role}
\IdrisFunction{tcpHandshakeDep} \IdrisBound{alice} \IdrisBound{bob} = \IdrisKeyword{do}
   \IdrisData{(_\IdrisData{,\IdrisBound{x\IdrisData{)}}}} <- \IdrisData{\IdrisBound{\IdrisBound{\IdrisBound{Send \IdrisBound{alice} \IdrisBound{bob} \IdrisType{(\IdrisType{Packet\IdrisType{,}}} \IdrisType{Nat\IdrisType{)}}}}}}
   \IdrisData{(_\IdrisData{,\IdrisBound{x'\IdrisData{,\IdrisBound{y\IdrisData{)}}}}}} <- \IdrisData{\IdrisBound{\IdrisBound{\IdrisBound{\IdrisBound{Send \IdrisBound{bob} \IdrisBound{alice} \IdrisType{(\IdrisType{Packet\IdrisType{,}}} \IdrisType{(\IdrisBound{xplus1}} \IdrisType{**} \IdrisBound{xplus1} \IdrisType{=} \IdrisData{S} \IdrisBound{x\IdrisType{)\IdrisType{,}}} \IdrisType{Nat\IdrisType{)}}}}}}}
   \IdrisData{Send} \IdrisBound{alice} \IdrisBound{bob} \IdrisType{(\IdrisType{Packet\IdrisType{,}}} \IdrisType{(\IdrisBound{samex}} \IdrisType{**} \IdrisBound{samex} \IdrisType{=} \IdrisBound{x'\IdrisType{)\IdrisType{,}}} \IdrisType{(\IdrisBound{yplus1}} \IdrisType{**} \IdrisBound{yplus1} \IdrisType{=} \IdrisData{S} \IdrisBound{y\IdrisType{)\IdrisType{)}}}
   \IdrisData{End}
\end{FigureVerbatim}


%% file: figure/naive/specification.error.tex
\begin{FigureVerbatim}
\IdrisFunction{tcpHandshakeErr} : (\IdrisBound{alice}, \IdrisBound{bob}, \IdrisBound{charlie} : \IdrisBound{role}) -> \IdrisType{Session} \IdrisType{()} \IdrisBound{role}
\IdrisFunction{tcpHandshakeErr} \IdrisBound{alice} \IdrisBound{bob} \IdrisBound{charlie} = \IdrisKeyword{do}
   \IdrisData{(_\IdrisData{,\IdrisBound{x\IdrisData{)}}}} <- \IdrisData{\IdrisBound{\IdrisBound{\IdrisBound{\IdrisBound{Send \IdrisBound{alice} \IdrisBound{bob} \IdrisType{(\IdrisType{Packet\IdrisType{,}}} \IdrisType{Nat\IdrisType{)}}}}}}}
   \IdrisData{(_\IdrisData{,\IdrisBound{x'\IdrisData{,\IdrisBound{y\IdrisData{)}}}}}} <- \IdrisData{\IdrisBound{\IdrisBound{\IdrisBound{\IdrisBound{\IdrisBound{\IdrisBound{Send \IdrisBound{bob} \IdrisBound{alice} \IdrisType{(\IdrisType{Packet\IdrisType{,}}} \IdrisType{(\IdrisBound{xplus1}} \IdrisType{**} \IdrisBound{xplus1} \IdrisType{=} \IdrisData{S} \IdrisBound{x\IdrisType{)\IdrisType{,}}} \IdrisType{Nat\IdrisType{)}}}}}}}}}
   \IdrisData{Send} \IdrisBound{charlie} \IdrisBound{alice} \IdrisType{(\IdrisType{Packet\IdrisType{,}}} \IdrisType{(\IdrisBound{xplus1}} \IdrisType{**} \IdrisBound{xplus1} \IdrisType{=} \IdrisData{S} \IdrisBound{x\IdrisType{)\IdrisType{,}}} \IdrisType{Nat\IdrisType{)}}
   \IdrisData{Send} \IdrisBound{alice} \IdrisBound{bob} \IdrisType{(\IdrisType{Packet\IdrisType{,}}} \IdrisType{(\IdrisBound{samex}} \IdrisType{**} \IdrisBound{samex} \IdrisType{=} \IdrisBound{x'\IdrisType{)\IdrisType{,}}} \IdrisType{(\IdrisBound{yplus1}} \IdrisType{**} \IdrisBound{yplus1} \IdrisType{=} \IdrisData{S} \IdrisBound{y\IdrisType{)\IdrisType{)}}}
   \IdrisData{End}
\end{FigureVerbatim}


%% file: content/language.tex
\section{A Language for Describing Sessions}\label{sec:global}

\begin{figure}[ht]
  \centering
  \input{./figure/language}
  \caption{Type definition for Sessions.}\label{fig:global:session}
\end{figure}

\Cref{fig:global:session} presents \texttt{Sessions}, our \ac{edsl} for specifying global session descriptions.
Messages are represented using variables that have an associated abstract state that keeps track of roles that have seen the message.
Central to the \acp{edsl} operation is a parameterised monad that manages the set of variables and their abstract state: The \emph{Knowledge Index}.

\begin{figure}[ht]
  \centering
  \begin{subfigure}{0.475\linewidth}
  \input{./figure/type-level}
  \caption{\label{fig:language:type-level:types}Definition.}
  \end{subfigure}
  \begin{subfigure}{0.475\linewidth}
  \input{./figure/operations}
  \caption{\label{fig:language:ki:operation}Operations.}
  \end{subfigure}
  \caption{\label{fig:language:ki}Knowledge Index Definition and Helpers.}
\end{figure}

\paragraph{Knowledge Index}
\Cref{fig:language:type-level:types} presents the various type-level data structures that make the knowledge index.
Roles within our \ac{edsl} are represented as an indexed data type tagged with a descriptive name.
The type \IdrisType{Role} is indexed by the type associated with roles to ensure that all roles are from the same family.
This representation of roles is a marked difference from the \naive{} implementation presented in \Cref{fig:intro:naive:def}.
Tagging of roles with a string value provides role comparison and reuse of existing predicates for list quantification.
Each message in a session description is represented by an indexed type \IdrisType{Var}, whose type is indexed by the type of the message being sent.
During type checking Idris' elaborator allows us to distinguish between different instances of \texttt{Var} based on their names and type-level values.
The knowledge index itself is a list of state items.
Items within our knowledge index associate a message type, with a variable, and a list of roles that are aware of the message.
Standard list quantifiers, such as \IdrisType{Any} and \IdrisType{All}, help reason about the knowledge context itself, and such quantifiers allow construction of operations acting on knowledge index instances---see \Cref{fig:language:ki:operation}.

\paragraph{Parameterised Type}
\texttt{Sessions} is a resource dependent \ac{edsl} in which the program's abstract state is the knowledge index.
The parameters of \IdrisType{Sessions} form a Hoare monad~\cite{Atkey2009pnc,Borgstrom2011vsp} where the resulting type-level state is dependent on the expression value:
\IdrisBound{type}---the expression's return type;
\IdrisBound{roleType}---the underlying type for roles;
\IdrisBound{participants}---the set of roles involved in the protocol;
\IdrisBound{old}---the knowledge index in which the expression must operate; and
\IdrisBound{new}---a function that computes the resulting knowledge index that is dependent upon the result associated with the expression.
A Let-binding allows us to sequence expressions, and bind message values to identifiers.
Each expression describes how the knowledge index is affected by the result of the expression.
When the result is not dependent on the expression value the function \IdrisFunction{const} (which drops its first parameter) allows one to state the new value.
When sequencing expressions, together with result dependent changes, the value of the knowledge index will change dynamically as we step through a session description.

\begin{figure}[ht]
  \centering
  \begin{subfigure}{0.475\linewidth}
    \input{./figure/KnowsData}
    \caption{\label{fig:language:predicates:knowsdata}Predicate stating that a role knows the message.}
  \end{subfigure}
  \begin{subfigure}{0.475\linewidth}
    \input{./figure/Learn}
    \caption{\label{fig:language:predicates:learn}Function to add role to abstract state.}
  \end{subfigure}
  \begin{subfigure}{1.0\linewidth}
    \input{./figure/AllKnow}
  \caption{\label{fig:language:predicates:operation}Predicate stating all the roles know the message.}
  \end{subfigure}
  \caption{\label{fig:language:predicates}Predicates reasoning on Items within the Knowlege Context, and operations that use these predicates.}
\end{figure}


\paragraph{Message Creation}
\IdrisData{NewMsg} and \IdrisData{NewDepMsg} are responsible for introducing message descriptions.
The constructor \IdrisData{NewMsg} will introduce a message of type \IdrisType{mType} created by \IdrisBound{creator} if \IdrisBound{creator} was in the global set of participants \IdrisBound{ps}.
A list quantifier provides this guarantee.
Creation of the message will extend the knowledge context with a new \IdrisType{Item} instance that populates the list of roles with \IdrisBound{creator}.
The constructor \IdrisData{NewDepMsg} will introduce a dependently typed message that depends on a previously seen value if said value is known to the message creator.
This proof is provided by \IdrisBound{prfKnows}, an instance of a list quantification that the predicate \IdrisType{KnowsData} holds for at least one element in the knowledge index.
\Cref{fig:language:predicates} presents the definition for \IdrisType{KnowsData} that states the given role is an element in the list of roles associated with the given message.
The parameter \IdrisBound{pred} is a function that constructs the type of the message to be sent, with the actual returned message type being a dependent pair that details the value associated with \IdrisBound{dep} that is passed to \IdrisBound{pred}.

\paragraph{Sending Messages}
The expression \IdrisData{Send} allows one to specify the type of message that is to be sent and which roles are sending and receiving the message.
As with all the other expressions evidence must be given to ensure expression correctness.
We require that both the sender and receiver are elements of \IdrisBound{ps}, and that the sender knows the message to be sent---\cf{} predicates associated with dependent message creation.
If a send is successfully described the knowledge index will be updated to add \IdrisBound{receiver} to the list of roles that know about \IdrisBound{msg} using the function \IdrisFunction{update}.

\paragraph{Accessing Values}
By design \IdrisType{Sessions} forbids message descriptions to be bound to a value explicitly, as not all participants will know the value.
However, if all participants in the protocol are aware of the message, then we can depend on the value.
The expression \IdrisData{Read} facilitates this.
The predicate \IdrisType{AllKnow} (presented in \Cref{fig:language:predicates}) constructs an inductive proof that each role in the session has knowledge of the message.
\Cref{sec:example} presents an example of \IdrisData{Read} in action.

\paragraph{Recursion}
Not all sessions comprise of a linear sequence of actions, interactions between roles may repeat.
\IdrisData{Rec} and \IdrisData{Call} allow for recursively calling an already described session, and calling an external session.
These functions are restricted such that the knowledge of the called session must begin and end with an empty context \ie{} no knowledge is learned about the supplied session.
A further restriction is on the set of participants respective to the calling function.
For \IdrisData{Rec} the caller and callee must have the same set, while a call to \IdrisData{Call} must have overlapping sets.
Here \IdrisType{Overlapping} is a thinning~\cite{Allais:2018:TSS:3243631.3236785} that ensures elements of \IdrisBound{ps} appear in order within \IdrisBound{ss}.
A thinning allows for structures to be weakened respective to some decision procedure~\cite{Chapman2009phdthesis,DBLP:conf/ctcs/AltenkirchHS95}.

\begin{figure}[ht]
  \centering
  \input{./figure/send}
  \caption{\label{fig:global:session-synonym}Example API for \IdrisData{Send}.}
\end{figure}

\paragraph{A Clean API}
Separately, we provide a high-level API for protocol designers to use when specifying protocols.
Without this API the proofs required for each expression have to be explicitly presented.
We can use Idris' auto implicit feature to automatically construct the proofs associated with an expression.
Idris' compiler will attempt to search and combine values, found in the expressions context, that together satisfy the type of the presented predicate.
If a suitable value cannot be constructed the expression will fail to type-check.
\Cref{fig:global:session-synonym} demonstrates this approach for the \IdrisData{Send} expression.
Further, we can provide a function (\IdrisFunction{Session}) to describe the expected initial and final state of the knowledge index together with the final expression type such that the only expression that will satisfy the end conditions would be the \IdrisData{End} expression.

%% file: figure/language.tex
\begin{FigureVerbatim}
\IdrisKeyword{data} \IdrisType{Sessions} : (\IdrisBound{type} : \IdrisType{Type})
             -> (\IdrisBound{roleType} : \IdrisType{Type})
             -> (\IdrisBound{participants} : \IdrisType{List} (\IdrisType{Role} \IdrisBound{roleType}))
             -> (\IdrisBound{old} : \IdrisType{List} (\IdrisType{Item} \IdrisBound{roleType}))
             -> (\IdrisBound{new} : \IdrisBound{type} -> \IdrisType{List} (\IdrisType{Item} \IdrisBound{roleType}))
             -> \IdrisType{Type} \IdrisKeyword{where}
  \IdrisData{Let} : (\IdrisBound{this} : \IdrisType{Sessions} \IdrisBound{a} \IdrisBound{roleType} \IdrisBound{ps} \IdrisBound{old} \IdrisBound{ctxt_fn})
     -> (\IdrisBound{into} : (\IdrisBound{value} : \IdrisBound{a}) -> \IdrisType{Sessions} \IdrisBound{b} \IdrisBound{roleType} \IdrisBound{ps} (\IdrisBound{ctxt_fn} \IdrisBound{value}) \IdrisBound{ctxt_fn_new})
     -> \IdrisType{Sessions} \IdrisBound{b} \IdrisBound{roleType} \IdrisBound{ps} \IdrisBound{old} \IdrisBound{ctxt_fn_new}

  \IdrisData{NewMsg} : (\IdrisBound{creator} : \IdrisType{Role} \IdrisBound{roleType}) -> (\IdrisBound{mType} : \IdrisType{Type}) -> (\IdrisBound{prfInvolved} : \IdrisType{Elem} \IdrisBound{creator} \IdrisBound{ps})
        -> \IdrisType{Sessions} (\IdrisType{Var} (\IdrisData{MSG} \IdrisBound{mType})) \IdrisBound{roleType} \IdrisBound{ps} \IdrisBound{old}
                   (\textbackslash{}\IdrisBound{lbl} => \IdrisData{MkItem} (\IdrisData{MSG} \IdrisBound{mType}) \IdrisBound{lbl} (\IdrisData{[\IdrisBound{creator\IdrisData{]}}}) \IdrisData{::} \IdrisBound{old})

  \IdrisData{NewDepMsg} : (\IdrisBound{creator} : \IdrisType{Role} \IdrisBound{roleType}) -> (\IdrisBound{dep} : \IdrisType{Var} (\IdrisData{MSG} \IdrisBound{mType}))
           -> (\IdrisBound{prfKnows} : \IdrisFunction{Any} (\IdrisType{KnowsData} \IdrisBound{creator} \IdrisBound{dep}) \IdrisBound{old}) -> (\IdrisBound{pred} : \IdrisBound{mType} -> \IdrisType{Type})
           -> \IdrisType{Sessions} (\IdrisType{Var} (\IdrisData{MSG} \IdrisType{(\IdrisBound{x}} \IdrisType{**} \IdrisBound{pred} \IdrisBound{x\IdrisType{)}})) \IdrisBound{roleType} \IdrisBound{ps} \IdrisBound{old}
                      (\textbackslash{}\IdrisBound{lbl} => \IdrisData{MkItem} (\IdrisData{MSG} \IdrisType{(\IdrisBound{x}} \IdrisType{**} \IdrisBound{pred} \IdrisBound{x\IdrisType{)}}) \IdrisBound{lbl} (\IdrisData{[\IdrisBound{creator\IdrisData{]}}}) \IdrisData{::} \IdrisBound{old})

  \IdrisData{Send} : (\IdrisBound{sender}, \IdrisBound{receiver} : \IdrisType{Role} \IdrisBound{roleType}) -> (\IdrisBound{msg} : \IdrisType{Var} (\IdrisData{MSG} \IdrisBound{mType}))
      -> (\IdrisBound{senderInvolved}   : \IdrisType{Elem} \IdrisBound{sender} \IdrisBound{ps}) -> (\IdrisBound{receiverInvolved} : \IdrisType{Elem} \IdrisBound{receiver} \IdrisBound{ps})
      -> (\IdrisBound{prf} : \IdrisFunction{Any} (\IdrisType{KnowsData} \IdrisBound{sender} \IdrisBound{msg}) \IdrisBound{old})
      -> \IdrisType{Sessions} \IdrisType{()} \IdrisBound{roleType} \IdrisBound{ps} \IdrisBound{old}
                 (\IdrisFunction{const} $ \IdrisFunction{update} \IdrisBound{old} \IdrisBound{prf} (\IdrisFunction{Learn} \IdrisBound{receiver}))

  \IdrisData{Rec} : \IdrisFunction{Inf} (\IdrisType{Sessions} \IdrisType{()} \IdrisBound{roleType} \IdrisBound{ps} \IdrisData{Nil} (\IdrisFunction{const} \IdrisData{Nil}))
      -> \IdrisType{Sessions} \IdrisType{()} \IdrisBound{roleType} \IdrisBound{ps} \IdrisBound{ctxt} (\IdrisFunction{const} \IdrisData{Nil})

  \IdrisData{Call} : \IdrisType{Sessions} \IdrisType{()} \IdrisBound{roleType} \IdrisBound{ss} \IdrisData{Nil} (\IdrisFunction{const} \IdrisData{Nil}) -> (\IdrisBound{prf} : \IdrisType{Overlapping} \IdrisBound{ps} \IdrisBound{ss})
      -> \IdrisType{Sessions} \IdrisType{()} \IdrisBound{roleType} \IdrisBound{ps} \IdrisBound{ctxt} (\IdrisFunction{const} \IdrisBound{ctxt})

  \IdrisData{End} : \IdrisType{Sessions} \IdrisType{()} \IdrisBound{roleType} \IdrisBound{ps} \IdrisBound{old} (\IdrisFunction{const} \IdrisData{Nil})

  \IdrisData{Read} : (\IdrisBound{msg} : \IdrisType{Var} (\IdrisData{MSG} \IdrisBound{mType})) -> (\IdrisBound{prf} : \IdrisFunction{Any} (\IdrisType{AllKnow} \IdrisBound{ps} \IdrisBound{msg}) \IdrisBound{old})
      -> \IdrisType{Sessions} \IdrisBound{mType} \IdrisBound{roleType} \IdrisBound{ps} \IdrisBound{old} (\IdrisFunction{const} \IdrisBound{old})
\end{FigureVerbatim}


%% file: figure/type-level.tex
\begin{FigureVerbatim}
\IdrisKeyword{data} \IdrisType{Role} : (\IdrisBound{type} : \IdrisType{Type}) \IdrisKeyword{where}
  \IdrisData{MkRole} : (\IdrisBound{tag} : \IdrisType{String}) -> \IdrisType{Role} \IdrisImplicit{type}

\IdrisKeyword{data} \IdrisType{\IdrisType{Ty}} = \IdrisData{MSG} \IdrisType{Type}

\IdrisKeyword{data} \IdrisType{Var} : (\IdrisBound{metaType} : \IdrisType{Ty}) -> \IdrisType{Type} \IdrisKeyword{where}
  \IdrisData{MkVar} : \IdrisType{Var} \IdrisBound{metatype}

\IdrisKeyword{data} \IdrisType{Item} : (\IdrisBound{type} : \IdrisType{Type}) -> \IdrisType{Type} \IdrisKeyword{where}
  \IdrisData{MkItem} : (\IdrisBound{msgtype} : \IdrisType{Ty})
        -> (\IdrisBound{label} : \IdrisType{Var} \IdrisBound{msgtype})
        -> (\IdrisBound{value} : \IdrisType{List} (\IdrisType{Role} \IdrisBound{type}))
        -> \IdrisType{Item} \IdrisBound{type}
\end{FigureVerbatim}


%% file: figure/operations.tex
\begin{FigureVerbatim}
\IdrisFunction{update} : (\IdrisBound{ctxt} : \IdrisType{List} (\IdrisType{Item} \IdrisBound{roleType}))
      -> (\IdrisBound{idx} : \IdrisFunction{Any} \IdrisBound{p} \IdrisBound{ctxt})
      -> (\IdrisBound{f} : (\IdrisBound{item} : \IdrisType{Item} \IdrisBound{roleType})
           -> (\IdrisBound{prf} : \IdrisBound{p} \IdrisBound{item})
           -> \IdrisType{Item} \IdrisBound{roleType})
      -> \IdrisType{List} (\IdrisType{Item} \IdrisBound{roleType})
\IdrisFunction{update} (\IdrisBound{x} \IdrisData{::} \IdrisBound{xs}) (\IdrisData{Here} \IdrisBound{y}) \IdrisBound{f} = \IdrisBound{f} \IdrisBound{x} \IdrisBound{y} \IdrisData{::} \IdrisBound{xs}
\IdrisFunction{update} (\IdrisBound{x} \IdrisData{::} \IdrisBound{xs}) (\IdrisData{There} \IdrisBound{y}) \IdrisBound{f} =
        \IdrisBound{x} \IdrisData{::} \IdrisFunction{update} \IdrisBound{xs} \IdrisBound{y} \IdrisBound{f}
\end{FigureVerbatim}

%% file: figure/KnowsData.tex
\begin{FigureVerbatim}
\IdrisKeyword{data} \IdrisType{KnowsData} : (\IdrisBound{role} : \IdrisType{Role} \IdrisBound{roleType})
              -> (\IdrisBound{msg} : \IdrisType{Var} (\IdrisData{MSG} \IdrisBound{typeM}))
              -> (\IdrisBound{item} : \IdrisType{Item} \IdrisBound{roleType})
              -> \IdrisType{Type} \IdrisKeyword{where}
  \IdrisData{WhoKnows} : \IdrisType{Elem} \IdrisBound{r} \IdrisBound{rs}
         -> \IdrisType{KnowsData} \IdrisBound{r} \IdrisBound{l} (\IdrisData{MkItem} (\IdrisData{MSG} \IdrisBound{ty}) \IdrisBound{l} \IdrisBound{rs})
\end{FigureVerbatim}

%% file: figure/Learn.tex
\begin{FigureVerbatim}
\IdrisFunction{Learn} : \IdrisType{Role} \IdrisBound{roleType}
     -> (\IdrisBound{i} : \IdrisType{Item} \IdrisBound{roleType})
     -> (\IdrisBound{prf} : \IdrisType{KnowsData} \IdrisBound{s} \IdrisBound{l} \IdrisBound{i})
     -> \IdrisType{Item} \IdrisBound{roleType}
\IdrisFunction{Learn} \IdrisBound{r} (\IdrisData{MkItem} (\IdrisData{MSG} \IdrisBound{ty}) \IdrisBound{l} \IdrisBound{rs}) (\IdrisData{WhoKnows} \IdrisBound{prf}) =
  \IdrisData{MkItem} (\IdrisData{MSG} \IdrisBound{ty}) \IdrisBound{l} (\IdrisBound{r}\IdrisData{::}\IdrisBound{rs})
\end{FigureVerbatim}

%% file: figure/AllKnow.tex
\begin{FigureVerbatim}
\IdrisKeyword{data} \IdrisType{AllKnow} : (\IdrisBound{roles} : \IdrisType{List} (\IdrisType{Role} \IdrisBound{roleType}))
            -> (\IdrisBound{msg} : \IdrisType{Var} (\IdrisData{MSG} \IdrisBound{typeM}))
            -> (\IdrisBound{item} : \IdrisType{Item} \IdrisBound{roleType}) -> \IdrisType{Type} \IdrisKeyword{where}
  \IdrisData{LastToKnow} : (\IdrisBound{prfKnows} : \IdrisType{Elem} \IdrisBound{r} \IdrisBound{rs}) -> \IdrisType{AllKnow} \IdrisData{[}\IdrisBound{r}\IdrisData{]} \IdrisBound{l} (\IdrisData{MkItem} (\IdrisData{MSG} \IdrisBound{ty}) \IdrisBound{l} \IdrisBound{rs})

  \IdrisData{NextToKnow} : \IdrisType{Elem} \IdrisBound{r} \IdrisBound{rs}
            -> \IdrisType{AllKnow} \IdrisBound{rs'} \IdrisBound{lbl} (\IdrisData{MkItem} (\IdrisData{MSG} \IdrisBound{ty}) \IdrisBound{lbl} \IdrisBound{rs})
            -> \IdrisType{AllKnow} (\IdrisBound{r}\IdrisData{::}\IdrisBound{rs'}) \IdrisBound{lbl} (\IdrisData{MkItem} (\IdrisData{MSG} \IdrisBound{ty}) \IdrisBound{lbl} \IdrisBound{rs})

\end{FigureVerbatim}

%% file: figure/send.tex
\begin{FigureVerbatim}
\IdrisFunction{send} : (\IdrisBound{sender}, \IdrisBound{receiver} : \IdrisType{Role} \IdrisBound{roleType})
    -> (\IdrisBound{msg} : \IdrisType{Var} (\IdrisData{MSG} \IdrisBound{mType}))
    -> \{\IdrisKeyword{auto} \IdrisBound{prfSender} : \IdrisType{Elem} \IdrisBound{sender} \IdrisBound{ps}\}
    -> \{\IdrisKeyword{auto} \IdrisBound{prfReceiver} : \IdrisType{Elem} \IdrisBound{receiver} \IdrisBound{ps}\}
    -> \{\IdrisKeyword{auto} \IdrisBound{prf} : \IdrisType{Any} (\IdrisType{KnowsData} \IdrisBound{sender} \IdrisBound{msg}) \IdrisBound{ctxt_old}\}
    -> \IdrisType{Session} \IdrisType{()} \IdrisBound{roleType} \IdrisBound{ps} \IdrisBound{ctxt_old} (\IdrisFunction{const} $ \IdrisFunction{update} \IdrisBound{ctxt_old} \IdrisBound{prf} (\IdrisFunction{Learn} \IdrisBound{receiver}))
\IdrisFunction{send} \IdrisBound{s} \IdrisBound{r} \IdrisBound{msg} \{\IdrisBound{prfSender}\} \{\IdrisBound{prfReceiver}\} \{\IdrisBound{prf}\} =
    \IdrisData{Send} \IdrisBound{s} \IdrisBound{r} \IdrisBound{msg} \IdrisBound{prfSender} \IdrisBound{prfReceiver} \IdrisBound{prf}
\end{FigureVerbatim}

%% file: content/examples.tex
\section{Example Sessions}\label{sec:example}

This section considers the expressiveness of our \ac{edsl} by considering three protocol descriptions.
The first example presents the TCP Handshake that demonstrates how we can construct value dependent message descriptions.
The second example is a multi-modal server that demonstrates how we can compose protocol descriptions together and make interaction decisions based of message values.
The final example demonstrates how we can construct higher-order protocols.

\subsection{TCP Handshake}\label{sec:example:tcp}

\begin{figure}[ht]
  \centering
  \input{./figure/example/tcp}
  \caption{\label{fig:intro:tcp-dep}A Dependently Typed Global Session Description for the TCP Handshake.}
\end{figure}

\Cref{fig:intro:tcp-dep} illustrates how the TCP handshake (presented earlier in \Cref{sec:introduction}) can be specified using \texttt{Sessions}.
Central to operation of the handshake is the sending of two sequence numbers that have been correctly transformed.
Within this example, Alice sends a non-dependently typed message (\IdrisBound{m1}) to Bob that contains the initial packet and Alice's sequence number.
In response, Bob constructs a value dependent message (\IdrisBound{m2}) that depends on the content of \IdrisBound{m1}.
The anonymous function states that \IdrisBound{m2} must have a type in which the second postion in the tuple is the sequence number from \IdrisBound{m1} incremented by one.
Here the type \IdrisFunction{Next} is a type synonym for a dependent pair stating the transformation on the sequence number from \IdrisBound{m1}.
The final message sent by Alice (\IdrisBound{m3}) is dependent on \IdrisBound{m2}.
Alice must send a packet together with the incremented sequence number from \IdrisBound{m2}, and the second sequence number incremented by one.
Recall that the type of \IdrisBound{m2} is a dependent pair in which the message type is in the second position and the dependend upon value is in the first.
Therefore to access the underlying values we have to project into these pairs, and their contents, accordingly.
Use of anonymous functions here does complicate the session description.
Such complication can be resolved with named functions.

\subsection{Multi-Modal Server}\label{sec:example:server}

\begin{figure}[hbt]
  \centering
  \begin{subfigure}{\textwidth}
    \input{./figure/example/server-types}
    \caption{\label{fig:example:server:data}Message Types and Values.}
  \end{subfigure}
  \begin{subfigure}{0.475\textwidth}
    \input{./figure/example/server-math}
    \caption{\label{fig:example:server:math}A Maths Protocol.}
  \end{subfigure}
  \begin{subfigure}{0.475\textwidth}
    \input{./figure/example/server-echo}
    \caption{\label{fig:example:server:echo}An Echo Protocol.}
  \end{subfigure}
  \begin{subfigure}{\textwidth}
    \input{./figure/example/server-core}
    \caption{\label{fig:example:server:core}The Server Protocol.}
  \end{subfigure}
  \caption{\label{fig:example:server}A Global Session Type for a Server.}
\end{figure}

\Cref{fig:example:server} presents the \emph{complete} specification of a global session description for interacting with a server that offers simple arithmetic calculations, and an echo service.
\Cref{fig:example:server:data} presents the messages sent.
Within this example:
Alice represents the client;
Bob the server; and
Charlie a third-party for performing simple arithmetic.
\Cref{fig:example:server:core} presents the top-level interactions between Alice and Bob.
Alice sends a message of type \IdrisType{CMD} to Bob.
Once Bob has received the message all participants specified in the top-level protocol have seen the value.
This enables the \IdrisData{Read} expression to access the value, and allow case-splitting (analogous to offer and choice from \ac{mpst}) to change behaviour based on the message's value.
If the command was:
\IdrisData{Math} then we call the Maths protocol and loop---\Cref{fig:example:server:math};
\IdrisData{Echo} then we call the Echo protocol and loop---\Cref{fig:example:server:echo}; or
\IdrisData{Quit} then we end the interaction.

The \IdrisFunction{doEcho} protocol is a non-recursive implementation of \texttt{RFC862}/\texttt{RFC347}~\cite{rfc862,rfc347} in which Bob repeats the message Alice sent, back to Alice.
We ensure this repetition of values using a dependent message with a predicate to reason about the sent value.
Here \IdrisFunction{Literal} is a type-synonym for a dependent pair with an equality predicate.
Note the use of \IdrisFunction{Literal} to ensure that the welcome message is the literal string value given.
The \IdrisFunction{doMath} protocol allows Alice to send Bob simple arithmetic expressions (\IdrisType{MathsCMD}), that Bob sends to Charlie.
Bob can use Charlie to generate a response to send to Alice.

\subsection{Higher-Order Protocols}\label{sec:example:hoppy}

\begin{figure}[ht]
  \centering
  \input{./figure/example/hoppy}
  \caption{\label{fig:example:hoppy}A Higher-Order Protocol.}
\end{figure}

\Cref{fig:example:hoppy} presents a \emph{Higher-Order Protocol}, in which we treat session descriptions as \emph{first class} constructs.
Rather than explicitly calling a named description (\cf{}~\Cref{fig:example:server:core}) we pass in descriptions as parameters.
The predicate \IdrisType{Overlapping} ensures that the participants of \IdrisBound{body} (the called description) overlap with those specified in \IdrisFunction{HoppyServer}.
The description presented in \Cref{fig:example:hoppy} presents a decision procedure reminiscent of an authentication procedure.
Alice sends some message that is checked by Bob who responds with a decision, and the protocol's next steps are based on that decision.
This is not secure but nonetheless demonstrates the potential power of \texttt{Sessions}.
We can construct protocol descriptions that are composable and higher-order.


%% file: figure/example/tcp.tex
\begin{FigureVerbatim}
\IdrisFunction{tcpHandShakeDep} : \IdrisFunction{Session} \IdrisBound{roleType} \IdrisData{[}\IdrisFunction{Alice}\IdrisData{,} \IdrisFunction{Bob}\IdrisData{]}
\IdrisFunction{tcpHandShakeDep} = \IdrisKeyword{do}
  \IdrisBound{m1} <- \IdrisFunction{newMsg} \IdrisFunction{Alice} \IdrisType{(\IdrisType{Packet\IdrisType{,}}} \IdrisType{Nat\IdrisType{)}}
  \IdrisFunction{send} \IdrisFunction{Alice} \IdrisFunction{Bob} \IdrisBound{m1}
  \IdrisBound{m2} <- \IdrisFunction{newDepMsg} \IdrisFunction{Bob} \IdrisBound{m1} (\textbackslash{}\IdrisBound{v} => \IdrisType{(\IdrisType{Packet\IdrisType{,}}} \IdrisFunction{Next} (\IdrisFunction{snd} \IdrisBound{v})\IdrisType{,} \IdrisType{Nat\IdrisType{)}})
  \IdrisFunction{send} \IdrisFunction{Bob} \IdrisFunction{Alice} \IdrisBound{m2}
  \IdrisBound{m3} <- \IdrisFunction{newDepMsg} \IdrisFunction{Alice} \IdrisBound{m2} (\textbackslash{}\IdrisBound{v} => \IdrisType{(\IdrisType{Packet\IdrisType{,}}}
     (\IdrisFunction{Literal} $ \IdrisFunction{fst} $ \IdrisFunction{snd} $ (\IdrisFunction{snd} \IdrisBound{v}))\IdrisType{,} (\IdrisFunction{Next} $ (\IdrisFunction{snd} $ \IdrisFunction{snd} $ (\IdrisFunction{snd} \IdrisBound{v})))\IdrisType{)})
  \IdrisFunction{send} \IdrisFunction{Alice} \IdrisFunction{Bob} \IdrisBound{m3}
  \IdrisFunction{end}
\end{FigureVerbatim}


%% file: figure/example/server-types.tex
\begin{FigureVerbatim}
\IdrisKeyword{data} \IdrisType{\IdrisType{MathsCMD}} = \IdrisData{Add} \IdrisType{Nat} \IdrisType{Nat} | \IdrisData{Sub} \IdrisType{Nat} \IdrisType{Nat} | \IdrisData{Div} \IdrisType{Nat} \IdrisType{Nat} | \IdrisData{Mul} \IdrisType{Nat} \IdrisType{Nat}
\IdrisKeyword{data} \IdrisType{\IdrisType{CMD}} = \IdrisData{Maths} | \IdrisData{Echo} | \IdrisData{Quit}
\end{FigureVerbatim}

%% file: figure/example/server-math.tex
\begin{FigureVerbatim}
\IdrisFunction{doMaths} : \IdrisFunction{Session} \IdrisBound{roleType} \IdrisData{[}\IdrisFunction{Alice}\IdrisData{,} \IdrisFunction{Bob}\IdrisData{,} \IdrisFunction{Charlie}\IdrisData{]}
\IdrisFunction{doMaths} = \IdrisKeyword{do}
  \IdrisBound{m1} <- \IdrisFunction{newMsg} \IdrisFunction{Bob} (\IdrisFunction{Literal} \IdrisData{"Time for Maths!"})
  \IdrisFunction{send} \IdrisFunction{Bob} \IdrisFunction{Alice} \IdrisBound{m1}
  \IdrisBound{m2} <- \IdrisFunction{newMsg} \IdrisFunction{Alice} \IdrisType{MathsCMD}
  \IdrisFunction{send} \IdrisFunction{Alice} \IdrisFunction{Bob} \IdrisBound{m2}
  \IdrisFunction{send} \IdrisFunction{Bob} \IdrisFunction{Charlie} \IdrisBound{m2}
  \IdrisBound{m3} <- \IdrisFunction{newMsg} \IdrisFunction{Charlie} \IdrisType{Nat}
  \IdrisFunction{send} \IdrisFunction{Charlie} \IdrisFunction{Bob} \IdrisBound{m3}
  \IdrisFunction{send} \IdrisFunction{Bob} \IdrisFunction{Alice} \IdrisBound{m3}
  \IdrisFunction{end}
\end{FigureVerbatim}


%% file: figure/example/server-echo.tex
\begin{FigureVerbatim}
\IdrisFunction{doEcho} : \IdrisFunction{Session} \IdrisBound{roleType} \IdrisData{[}\IdrisFunction{Alice}\IdrisData{,} \IdrisFunction{Bob}\IdrisData{]}
\IdrisFunction{doEcho} = \IdrisKeyword{do}
  \IdrisBound{m1} <- \IdrisFunction{newMsg} \IdrisFunction{Bob} (\IdrisFunction{Literal} \IdrisData{"Time to Echo!"})
  \IdrisFunction{send} \IdrisFunction{Bob} \IdrisFunction{Alice} \IdrisBound{m1}
  \IdrisBound{m2} <- \IdrisFunction{newMsg} \IdrisFunction{Alice} \IdrisType{String}
  \IdrisFunction{send} \IdrisFunction{Alice} \IdrisFunction{Bob} \IdrisBound{m2}
  \IdrisBound{m3} <- \IdrisFunction{newDepMsg} \IdrisFunction{Bob} \IdrisBound{m2} (\IdrisFunction{Literal})
  \IdrisFunction{end}
\end{FigureVerbatim}


%% file: figure/example/server-core.tex
\begin{FigureVerbatim}
\IdrisFunction{myServer} : \IdrisFunction{Session} \IdrisBound{roleType} \IdrisData{[}\IdrisFunction{Alice}\IdrisData{,} \IdrisFunction{Bob}\IdrisData{]}
\IdrisFunction{myServer} = \IdrisKeyword{do}
  \IdrisBound{m1} <- \IdrisFunction{newMsg} \IdrisFunction{Alice} \IdrisType{CMD}
  \IdrisFunction{send} \IdrisFunction{Alice} \IdrisFunction{Bob} \IdrisBound{m1}
  \IdrisBound{m1val} <- \IdrisFunction{read} \IdrisBound{m1}
  \IdrisKeyword{case} \IdrisType{\IdrisBound{\IdrisBound{\IdrisBound{m1val}}}} \IdrisKeyword{of}
    \IdrisData{Maths} => \IdrisKeyword{do} \{\IdrisFunction{call} \IdrisFunction{doMaths}; \IdrisData{Rec} \IdrisFunction{myServer}\} \IdrisData{Echo} => \IdrisKeyword{do} \{\IdrisFunction{call} \IdrisFunction{doEcho}; \IdrisData{Rec} \IdrisFunction{myServer}\}; \IdrisData{Quit} => \IdrisFunction{end}\}
\end{FigureVerbatim}


%% file: figure/example/hoppy.tex
\begin{FigureVerbatim}
\IdrisFunction{HoppyServer} : \IdrisFunction{Session} \IdrisBound{roleType} \IdrisBound{ss}
           -> \{\IdrisKeyword{auto} \IdrisBound{prf} : \IdrisType{Overlapping} \IdrisData{[\IdrisFunction{Alice\IdrisData{,\IdrisFunction{Bob\IdrisData{]}}}}} \IdrisBound{ss}\}
           -> \IdrisFunction{Session} \IdrisBound{roleType} \IdrisData{[}\IdrisFunction{Alice}\IdrisData{,}\IdrisFunction{Bob}\IdrisData{]}
\IdrisFunction{HoppyServer} \IdrisBound{body} = \IdrisKeyword{do}
  \IdrisBound{m1} <- \IdrisFunction{newMsg} \IdrisFunction{Bob} (\IdrisFunction{Literal} \IdrisData{"Who are you!"})
  \IdrisFunction{send} \IdrisFunction{Bob} \IdrisFunction{Alice} \IdrisBound{m1}
  \IdrisBound{m2} <- \IdrisFunction{newMsg} \IdrisFunction{Alice} \IdrisType{String}
  \IdrisFunction{send} \IdrisFunction{Alice} \IdrisFunction{Bob} \IdrisBound{m2}
  \IdrisBound{m3} <- \IdrisFunction{newMsg} \IdrisFunction{Bob} \IdrisType{Bool}
  \IdrisFunction{send} \IdrisFunction{Bob} \IdrisFunction{Alice} \IdrisBound{m3}
  \IdrisBound{res} <- \IdrisFunction{read} \IdrisBound{m3}
  \IdrisKeyword{case} \IdrisType{\IdrisBound{\IdrisBound{\IdrisBound{\IdrisBound{\IdrisBound{\IdrisBound{\IdrisBound{\IdrisBound{res}}}}}}}}} \IdrisKeyword{of} \{ \IdrisData{True} => \IdrisKeyword{do} \{\IdrisFunction{call} \IdrisBound{body}; \IdrisFunction{end}\}; \IdrisData{False} => \IdrisFunction{end}\}
\end{FigureVerbatim}


%% file: content/discussion.tex
\section{Discussion}\label{sec:discussion}

\texttt{Sessions} is an \ac{edsl} for describing global session descriptions within a dependently typed host language.
We introduce value dependent messages and reason about messages within a paramterised monad.
Construction of \texttt{Sessions} as a \ac{edsl} allows session descriptions to be first class, composable, and computable.
Overriding \texttt{Do} notation facilitates use of Idris' control structures to describe decisions.
Here choice differs from branching/selection as traditionally seen in session type implementations.
\texttt{Sessions} supports value based choice using pattern-matching on message values or constant values.

\acp{mpst} can provide guarantees towards several protocol properties.
Namely, session fidelity, communication safety, liveness, and progress~\cite{DBLP:journals/jacm/HondaYC16,DBLP:conf/fase/HuY17}.
Many of these properties are for \emph{complete systems}, \texttt{Sessions} describes global descriptions only.
Existing work has shown a correct-by-construction approach to linking session descriptions to implementations within a dependently typed language~\cite{DBLP:journals/aghcs/Brady17}.
Global descriptions are projected to compute the local type for the viewpoint of a protocol participant as a continuation.
The type of the implementation is indexed by the continuation to ensure there is an intrinsic link between the specified global session description and its implementation.
Thus providing \emph{communication safety} and \emph{session fidelity}.
We are currently investigating how to build a similar framework that includes our value dependent session descriptions.

Further, Idris is a total language that checks for program termination and coverage of pattern matching clauses.
Thus, our global session descriptions are checked for termination and coverage in the same way as regular Idris programs.
We believe that implementing local types and implementations within a total language will help to provide guarantees towards \emph{liveness} and \emph{progress}.


%% file: content/related.tex
\section{Related Work}\label{sec:related-work}

There are many implementations of Session Types available\furl{http://simonjf.com/2016/05/28/session-type-implementations.html}.
Generally, there are three approaches to implementation either:
as a \ac{dsl};
as an \ac{edsl}; or
as a language extension.

\paragraph{\ac{dsl}}
Scribble is a \ac{dsl} for describing \acp{mpst}~\cite{Yoshida2014spl}.
The \ac{dsl} is limited in describing non-value dependent message exchanges.
There are various existing runtimes that generate code from these descriptions~\cite{Ng2012Msc,Kouzapas2016tpm,Castro:2019:DPU:3302515.3290342}.
As \texttt{Sessions} is an \ac{edsl} the resulting specifications cannot not be directly resused by other projects\footnote{Idris does, however, support multiple code generation targets, however, embedding compiled Idris code into other code projects is non-trivial.}.
However, the library can be extended to generate Scribble specifications directly from the \texttt{Session} specification.
StMungo is a Java oriented tool for generating local types from Scribble descriptions~\cite{Kouzapas2016tpm}.

\paragraph{Scribble-Refined}
Recent work~\cite{DBLP:conf/cc/NeykovaHYA18} has extended Scribble specifically for F\# to leverage the language's support for refinement types~ \cite{DBLP:conf/pldi/FreemanP91} and type providers~\cite{DBLP:conf/pldi/PetricekGS16}.
The authors present an expressive refinement language to specify boolean refinements related to messages being sent.
\texttt{Sessions} complements this work by showing an alternative realisation to reasoning about value messages.

\paragraph{\ac{edsl}}
\acp{edsl} have been realised for:
Haskell~\cite{Sackman2008sth,Orchard2016ess,session-haskell}; Go~\cite{DBLP:conf/popl/LangeNTY17}; and
Rust~\cite{session-rust}.
These approaches follow our approach, however, the host language chosen is not as expressive in describing global session description and thus do not allow for reasoning on message descriptions.

\paragraph{Language Extension}
Not all implementations follow the \ac{dsl} approach.
Others have extended existing programming languages to embedded \ac{mpst} directly within the language.
For example, Links and SIL~\cite{Fowler:2019:EAS:3302515.3290341,Lindley2017lfs,Balzer2015osp}.
However, these approaches require extending the language and compiler.
\texttt{Sessions} presents an opportunity to develop the specification in the same language as the implementation without major changes required to the language itself.

\paragraph{\ac{mpst} Theory}
Our work in using a dependently typed language to realise value dependent global session descriptions compliments existing theoretical work~\cite{DBLP:conf/fossacs/ToninhoY18,Toninho2016,DBLP:conf/concur/BocchiHTY10}.
\texttt{Sessions} achieves value-dependent session descriptions by maintaining a \emph{knowledge index}.
This compliments existing work~\cite{Toninho2016,DBLP:conf/cc/NeykovaHYA18} in which the authors define knowledge in much the same way.
However, we have embedded the knowledge index directly within the type-system of the global session description ensuring that our \ac{edsl} instances are correct-by-construction with respect to value dependencies.
A \emph{Design-By-Contract} approach has been taken to extend \ac{mpst} with message oriented assertions~\cite{DBLP:conf/concur/BocchiHTY10}.
We remark that the assertions associated with messages are comparable to the dependent pair construct in which the message (first position) must satisfy the predicate in the second position.
Our use of dependent pair's is different: the value in the first position presents evidence that the predicate (message) in the second position is valid.


%% file: content/conclusion.tex
\section{Conclusion}\label{sec:conclusion}

Type systems in modern programming languages are expressive enough to support \acp{edsl} describing session types.
With the additional expressivity given by dependent types, we have shown that dependently typed languages such as Idris provide a natural setting to further enhance the expressiveness of \acp{edsl} for describing global session descriptions.

\texttt{Sessions} demonstrates how we incorporate reasoning on value-dependencies between messages, and provide first-class global session descriptions.
Idris' auto implicit mechanism has proven useful in presenting correct-by-construction guarantees towards protocol design.
With this new setting for global session design we expect to be able to use Idris' type-system to verify additional correctness guarantees of our global descriptions.
Especially properties required by \emph{security} protocols~\cite{gordon2003authenticity,DBLP:conf/uss/Kaloper-Mersinjak15} where we also need to reason about the content of messages, and more importantly how messages are related.

\texttt{Sessions} is an \ac{edsl} for global session descriptions.
We wish to compliment our \ac{edsl} with a complete system for protocol design, implementation, and verification such that we provide a self-contained system within a single language.
Of importance will be how we can correctly project our descriptions to local types such that only correct local actions and knowledge are carried over.
